\documentclass[aps,prc,twocolumn,superscriptaddress,showpacs,floatfix]{revtex4}

\usepackage{graphicx}
\usepackage{dcolumn}
\usepackage{bm}
\usepackage{xspace}

\begin{document}

\newcommand{\ox}      {$^{16}$O$(p,2p)^{15}$N\xspace}
\newcommand{\sk}      {Super-Kamiokande\xspace}
\newcommand{\nukp}    {$p \to \bar{\nu} K^+$\xspace}
\newcommand{\tnu}     {$n \to \nu \nu \bar{\nu}$\xspace}
\newcommand{\resbras}  {15.6$\pm$1.3$^{+0.6}_{-1.0}$\%\xspace}
\newcommand{\resbra}  {15.6$\pm$1.3$^{+0.6}_{-1.0}$\% and 27.9$\pm$1.5$^{+3.4}_{-2.6}$\%\xspace}
\newcommand{\respro}  {3.1\% and 5.6\%\xspace}

\preprint{APS/123-QED}

\title{De-excitation $\gamma$-rays from the $s$-hole state in $^{15}$N
associated with proton decay in $^{16}$O}

\newcommand{\icrr}{\affiliation{Kamioka Observatory, Institute for Cosmic Ray Research, University of Tokyo, Kamioka, Gifu, 506-1205, Japan}}
\newcommand{\kyoto}{\affiliation{Department of Physics, Kyoto University, Kyoto 606-8502, Japan}}
\newcommand{\jasri}{\affiliation{Japan Synchrotron Radiation Research Institute, Sayo, Hyogo 679-5198, Japan}}
\newcommand{\konan}{\affiliation{Department of Physics, Konan University, Kobe 658-8501, Japan}}
\newcommand{\rcnp}{\affiliation{Research Center for Nuclear Physics, Osaka University, Ibaraki, Osaka 567-0047, Japan}}
\newcommand{\kansai}{\affiliation{Kansai Photon Science Institute, Japan Atomic Energy Agency, Kizu, Kyoto 619-215, Japan}}
\newcommand{\kangaku}{\affiliation{Laboratory of Physics, Kanto Gakuin University, Yokohama 236-0032, Japan}}
\newcommand{\jaeri}{\altaffiliation{Advanced Science Research Center, JAERI, Tokai, Ibaraki 319-1195, Japan}} 

\newcommand{\stonybrooknow}{\altaffiliation{Present address: Department of Physics and Astronomy, State University of New York, Stony Brook, NY 11794-3800, USA}}
\newcommand{\kyotonow}{\altaffiliation{Present address: Department of Physics, Kyoto University, Kyoto 606-8502, Japan}}
\newcommand{\keknow}{\altaffiliation{Present address: High Energy Accelerator Research Organization (KEK), Tsukuba, Ibaraki, 305-0801, Japan}}
\newcommand{\jncnow}{\altaffiliation{Present address: Japan Nuclear Cycle Development Institute, Tokai, Ibaraki 319-1194, Japan}}
\newcommand{\lnsnow}{\altaffiliation{Present address: Laboratory of Nuclear Science, Tohoku University, Sendai 982-0826, Japan}}
\newcommand{\cyricnow}{\altaffiliation{Present address: Cyclotron and Radioisotope Center, Tohoku University, Sendai 980-8578, Japan}}
\newcommand{\nagoyanow}{\altaffiliation{Present address: Solar-Terrestrial Environment Laboratory, Nagoya University, Nagoya 464-8601, Japan}}
\newcommand{\cnsnow}{\altaffiliation{Present address: Center for Nuclear Study, University of Tokyo, Bunkyo, Tokyo 113-0033, Japan}}
\newcommand{\wakayamanow}{\altaffiliation{Present address: Wakayama Medical University, Wakayama 641-8509, Japan}}
\newcommand{\rikennow}{\altaffiliation{Present address: RI Beam Science Laboratory, RIKEN, Wako, Saitama 351-0198, Japan}}
\newcommand{\titnow}{\altaffiliation{Present address: Department of Physics, Tokyo Institute of Technology, Meguro, Tokyo 152-8520, Japan}}
\newcommand{\kyushunow}{\altaffiliation{Present address: Research and Development Center for Higher Education, Kyushu University, Fukuoka 810-8560, Japan}}
\newcommand{\rcnpnow}{\altaffiliation{Present address: Research Center for Nuclear Physics, Osaka University, Ibaraki, Osaka 567-0047, Japan}}
\newcommand{\michigannow}{\altaffiliation{Present address: National Superconducting Cyclotron Laboratory, Michigan State University, East Lansing, Michigan 48824, USA}}

\author{K.~Kobayashi}\stonybrooknow\icrr
\author{H.~Akimune}\konan
\author{H.~Ejiri}\jasri\rcnp
\author{H.~Fujimura}\kyotonow\rcnp
\author{M.~Fujiwara}\rcnp\kansai
\author{K.~Hara}\keknow\rcnp
\author{K.Y.~Hara}\jncnow\konan
\author{T.~Ishikawa}\lnsnow\kyoto
\author{M.~Itoh}\cyricnow\rcnp
\author{Y.~Itow}\nagoyanow\icrr
\author{T.~Kawabata}\cnsnow\kyoto
\author{M.~Nakamura}\wakayamanow\kyoto
\author{H.~Sakaguchi}\kyoto
\author{Y.~Sakemi}\rcnp
\author{M.~Shiozawa}\icrr
\author{H.~Takeda}\rikennow\kyoto
\author{Y.~Totsuka}\keknow\icrr
\author{H.~Toyokawa}\jasri
\author{M.~Uchida}\titnow\rcnp
\author{T.~Yamada}\kangaku
\author{Y.~Yasuda}\kyoto
\author{H.P.~Yoshida}\kyushunow\rcnp
\author{M.~Yosoi}\rcnpnow\kyoto
\author{R.G.T.~Zegers}\michigannow\rcnp
       
\date{\today}

\begin{abstract}
We have measured de-excitation $\gamma$-rays from the $s$-hole state 
in $^{15}$N produced via the \ox reaction in relation to the study of the 
nucleon decay and the neutrino neutral-current interaction in water 
Cherenkov detectors.
In the excitation-energy region of the $s$-hole state between 16~MeV and 
40~MeV in $^{15}$N, the branching ratio of emitting $\gamma$-rays with 
the energies at more-than-6~MeV are found to be \resbras.
Taking into account the spectroscopic factor of the $s$-hole state,
the total emission probability is found to be 3.1\%. 
This is about 1/10 compared with the emission probability of the 6.32~MeV 
$\gamma$-ray from the 3/2$^-$~$p$-hole state in $^{15}$N.
Moreover, we searched for a 15.1~MeV $\gamma$-ray from the $^{12}$C~$1^+$ 
state which may be populated after the particle-decay of the $s$-hole
state in $^{15}$N.
Such a high energy $\gamma$-ray from the hole state would provide a new 
method to search for mode-independent nucleon decay even if the emission 
probability is small.
No significant signal is found within a statistical uncertainty.
\end{abstract}

\maketitle

\section{\label{sec:motiv}Introduction}
In the recent searches of proton decay and in the studies of neutrino 
oscillations, water Cherenkov detectors are used, such as \sk~\cite{sk}.
Since in such detectors, proton decays or neutrino interactions happen 
mostly in $^{16}$O, it is important to study the effect of the nuclear 
de-excitation process on the Cherenkov light detection in more detail.

When a proton decays in one of the inner-shell orbits of $^{16}$O,
the residual $^{15}$N nucleus remains in an excited state with a proton-hole.
This state quickly de-excites by emitting $\gamma$-rays with a certain 
probability.
In the decay search involving the channel \nukp~\cite{nuk}, which is the 
dominant process in many supersymmetric grand unification 
models~\cite{susygut1,susygut2,susygut3,susygut4,susygut5}, the 
de-excitation $\gamma$-rays would be very useful to reduce backgrounds.
Since the $K^+$ momentum is below the Cherenkov threshold, 
the $K^+$ is invisible. 
Therefore, the $\mu^+$ signal stemming from the $K^+$ decay can be 
separated from the prompt $\gamma$-ray signal.
On the other hand, most backgrounds produced by events such as
$\nu N \rightarrow \mu N^{\prime}\gamma$ have no time difference 
between the $\gamma$-ray and $\mu$ signals and can be eliminated
when the $\gamma$-ray energy ($E_{\gamma}$) is in the detectable energy 
range of the \sk.

The de-excitation $\gamma$-ray is also useful for the study of the neutrino
neutral-current interaction via the quasi-free knock-out process,
e.g. with a 1~kt water Cherenkov detector~\cite{k2k1kt} in the experiment
of the long base-line neutrino oscillation from KEK to \sk (K2K).
An overall neutrino flux independent of neutrino flavor can be estimated 
from the measurement of the neutral-current interaction.
Because recoil energies of protons are mostly below the Cherenkov threshold,
Cherenkov light from the de-excitation $\gamma$-rays can be clearly observed.
It is then possible to tag the neutral-current interaction
with the $\gamma$-rays.

The de-excitation process of hole states in $^{15}$N following proton
decay in $^{16}$O was discussed by Ejiri~\cite{ejiri}.
He summarized de-excitation modes and their branching ratios, and estimated
the emission probabilities of the de-excitation $\gamma$-rays.
For a hole state $k$ which de-excites only by $\gamma$~decay,
he estimated the $\gamma$-ray emission probability ($P(k)$) by
using the spectroscopic factor ($S_p(k)$) as $P(k)$=$S_p(k)$/8,
where 8 is the number of protons in $^{16}$O.
The 6.32~MeV $\gamma$-ray from the $3/2^-$ $p$-hole state to the ground
state (g.s.) in $^{15}$N, which is in the detectable energy range 
of the \sk detector, is the most probable one because the state has the 
largest spectroscopic factor and its excitation energy is below the 
particle emission threshold.
Using the spectroscopic factor from the ($d$,$^{3}$He) data~\cite{sp6mev}
(see Table~\ref{tab:comp}), Ejiri predicted that the emission probability 
of the 6.32~MeV $\gamma$-ray is 41\% 
($P(6.32{\rm \ MeV\ } p_{3/2^-})=S_p(6.32{\rm \ MeV\ }p_{3/2^-})/8=3.3/8$).
He also estimated the $\gamma$-ray emission probabilities from other $p$-hole
and $s$-hole states to be several percent, respectively.
However, the emission probabilities from the $s$-hole state around 25~MeV
excitation are uncertain.
The $s$-hole state de-excites mostly by particle emissions, because 
the excitation energy exceeds the particle emission threshold (10.21~MeV for 
proton decay and 10.83~MeV for neutron decay).
If the residual nucleus remains in an excited state, secondary $\gamma$-rays 
are emitted. 
Ejiri statistically evaluated the emission probabilities
by $P(k)=S_p(s{\rm -state})/8*S_p^{\prime}(k)/N_p*b(N)$, where
$S_p^{\prime}(k)$ is the spectroscopic factor of the excited state $k$ 
in the daughter nucleus, $N_p$ (=11) is the number of $p$-shell nucleons in 
$^{15}$N, and $b(N)$ is the escape ratio of the nucleon.
The excited states most likely to be populated were the 7.01~MeV $2^+$ state
in $^{14}$C and the 7.03~MeV $2^+$ state in $^{14}$N each with $P(k)=2\%$.

However, a recent study~\cite{yosoi} showed that the particle-decay data
of deep-hole states in light nuclei cannot be reproduced in a
statistical-model calculation.
The probability to remain the secondary excited states after particle 
emissions of the $s$-hole state in $^{15}$N was found to be high.
The results suggested that the direct two-body decay process from the
door-way $s$-hole state predicted by the theoretical calculations based on a
SU(3)-model~\cite{yamada} and a shell model~\cite{yamada2} considerably occurred.
Thus, from a nuclear-structure point of view, it is also interesting to 
measure such secondary de-excitation $\gamma$-rays.

After the work by Ejiri~\cite{ejiri}, the spectroscopic factors of $p$-hole
state have been investigated in an $^{16}$O$(e,e^{\prime}p)^{15}$N 
experiment~\cite{leus}.
The result suggests that the spectroscopic factors in the $p$-shell orbits
are substantially lower than the sum-rule limit ($S_p(p_{3/2})=4$) in the 
independent-particle shell-model; the spectroscopic factor of the 6.32~MeV 
$p$-state is smaller than that given by Ref.~\cite{ejiri} by 30\% as shown in 
Table~\ref{tab:comp}.
The difference is mainly due to a large theoretical uncertainty associated
with the reaction mechanism. 
The spectroscopic factor derived from transfer reactions strongly
depends on the delicate details of the distorted wave Born 
approximation~\cite{d3here}.
This result indicates that the direct measurement of $\gamma$-rays from 
the $s$-hole state is important.
\begin{center}
\begin{table}
\caption{\label{tab:comp} Comparison of spectroscopic factors of the dominant $p$-hole states in $^{15}$N. The data are taken from Ref.~\cite{ejiri} and \cite{leus}. }
\begin{tabular}{cc@{\hspace{0.5cm}}c@{\hspace{0.3cm}}c}
\hline
\hline
Energy (MeV) & J$^{\pi}$ & Ref.~\cite{ejiri} & Ref.~\cite{leus}\\
\hline
0       & 1/2$^-$ & 2    & 1.26 \\
5.27    & 5/2$^+$ & ---  & 0.11 \\
6.32    & 3/2$^-$ & 3.3  & 2.35 \\
9.93    & 3/2$^-$ & 0.26 & 0.13 \\
\hline
\hline
\end{tabular}
\end{table}
\end{center}

The nucleon-hole state is created via any modes of nucleon decay.
The de-excitation $\gamma$-ray is a mode-independent signal of the 
nucleon decay.
If the decay particles do not emit Cherenkov light (for example in \tnu decay),
the only detectable signal is Cherenkov light from the $\gamma$-ray. 
Since in the energy region below 10~MeV the signal in the \sk is 
contaminated by strong backgrounds from radioactivities, $\gamma$-rays with
$E_{\gamma}>$10~MeV are useful to tag the nucleon decay.
Though the energies of most $\gamma$-rays from the $s$-hole state in $^{15}$N
are expected to be less than 10~MeV, two different $\gamma$-rays 
over 10~MeV can be emitted.
One is the direct $\gamma$-ray with $E_{\gamma}$=20--30~MeV from the unbound 
$s$-hole state.
However, it was estimated that the emission probability is quite 
small~\cite{n3nukam}.
The other is a unique 15.1~MeV $\gamma$-ray which can be emitted mainly 
from the T=1~$1^+$ state in $^{12}$C after $t$~decay, $n$+$d$~decay, or
2$n$+$p$~decay from the $s$-hole state in $^{15}$N.
This $\gamma$-ray is useful to tag the mode-independent nucleon decay 
with small backgrounds in \sk.

In this paper, we present the first measurement of the de-excitation
$\gamma$-rays with $E_{\gamma}<$18~MeV from the $s$-hole state in
$^{15}$N produced via the \ox reaction.
These $\gamma$-rays are useful for the proton decay search and the study
of neutrino neutral-current interaction.
Especially, in the dominant method of proton decay search via \nukp 
in \sk, the partial lifetime of proton is estimated to be proportional 
to the emission probability of the de-excitation $\gamma$-ray.
Therefore, this measurement is important to obtain the precise emission
probability.
It is also important for determining the proton decay sensitivity in future
experiments~\cite{uno,hk}.
In terms of the 15.1~MeV $\gamma$-ray from the $^{12}$C~1$^+$ state,
even if the emission probability of this $\gamma$-ray is down to 
0.01\%, it would provide a new method to search for mode-independent 
nucleon decay above the current decay limit obtained by SNO~\cite{sno}.
Because the proton $s$-hole state in $^{15}$N and neutron $s$-hole state 
in $^{15}$O are isospin-symmetric, the $^{12}$C~1$^+$ state is expected
to be produced similarly.
The present measurement is also useful for neutron decay search, 
such as \tnu mode.

\section{\label{sec:exp}Experiment}
The experiment was carried out at the Research Center for
Nuclear Physics (RCNP), Osaka University using
392~MeV proton beam from the cyclotron facility~\cite{beam}.
The proton beam impinged on a H$_2$O ice target~\cite{kawa} in order
to generate the \ox reaction.
Two protons were detected in coincidence in the dual magnetic spectrometer 
system, consisting of Grand Raiden (GR)~\cite{gr} and Large Acceptance 
Spectrometer (LAS)~\cite{las}.
Both GR and LAS consisted of magnets, two multi-wire drift chambers (MWDCs),
and two plastic scintillators.
The laboratory angles of GR (25.5$^{\circ}$) and LAS (51.0$^{\circ}$)
relative to the beam and their magnetic fields were chosen to maximize 
the cross section of the ($p,2p$) reaction leading to the $s$-hole state 
when the recoil momentum of the residual nucleus is zero. 
The $\Delta E$ signals of the plastic scintillators were used for particle
identification, and were also used as trigger signals.
The solid angles of the spectrometers were 4.3~msr for GR and 20.0~msr for 
LAS, respectively.
The momenta and scattering angles of protons were determined by a ray-tracing
technique with MWDCs.
The experimental details for the production of $s$-hole states in light
nuclei via the ($p,2p$) reaction were reported in Refs.~\cite{yosoi,yosoi2}.

The de-excitation $\gamma$-rays were measured by three arrays of
NaI (Tl) detectors in coincidence with two protons.
Each detector array consisted of nine NaI scintillators with sizes of 
5.1~cm $\times$ 5.1~cm $\times$ 15.2~cm.
Three arrays were mounted at distances of 21.7~cm, 21.7~cm, and 26.8~cm 
from the target and at scattering angles of 117.4$^{\circ}$, 117.4$^{\circ}$,
and 147.5$^{\circ}$ with respect to the beam direction, respectively.
The total solid angle covered for the $\gamma$-ray detection was about 1.3~sr.
The scintillating light was measured by photomultiplier tubes (PMTs) and 
the pulse height and timing were recorded.

\section{\label{sec:ana}Analysis}
To identify the hole state, the excitation energy ($E_x$) is evaluated
from the energies of two emerging protons measured by GR and LAS as follows:
\begin{equation}
{\bf k}_3 = {\bf k}_0 - {\bf k}_1 - {\bf k}_2
\label{eq:mom}
\end{equation}
\begin{equation}
E_x - Q = T_0 - (T_1 + T_2 + T_3),
\label{eq:ene}
\end{equation}
where ${\bf k}_i$ and $T_i$ are the momenta and kinetic energies of the
incident proton ($i$=0), the emerging protons ($i$=1,2), and the recoiling
residual nucleus ($i$=3), respectively. 
The $Q$-value is given by $Q=M-(m_p+M^{\prime})$, where $M$, $M^{\prime}$, and
$m_p$ are the masses of the target $^{16}$O, residual nucleus $^{15}$N, and
proton, respectively.
\begin{figure*}[tbp]
\includegraphics[width=14.cm]{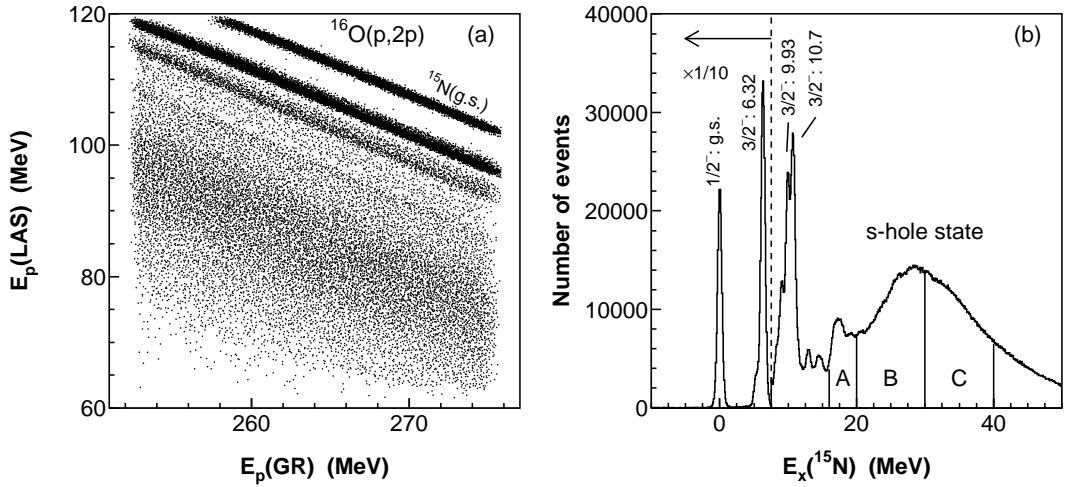}
\caption{\label{fig:miss} (a) The two-dimensional scatter plot of the kinetic energies of two protons ($E_p$) measured by GR and LAS in the \ox reaction. (b) The excitation energy spectrum of $^{15}$N induced by the \ox reaction. The region shown by an arrow is scaled by 1/10. In the analyses of the $\gamma$-ray from the $s$-hole state, we use the coincident data with two protons in the regions A ($E_x$=16--20~MeV), B ($E_x$=20--30~MeV), and C ($E_x$=30--40~MeV). }
\end{figure*}
Figure~\ref{fig:miss} shows: (a) the two-dimensional scatter plot of the
kinetic energies of two protons measured by GR and LAS, (b) the excitation 
energy spectrum of $^{15}$N induced by the \ox reaction after the accidental 
coincidence background is subtracted.
The $s$-hole state is strongly excited in the higher excitation energy
region ($E_x \ge$ 16~MeV) and splits into a few sub-structures.
This structure agrees qualitatively with the result of recent shell-model 
calculation~\cite{yamada2}.
Two peaks at 0.0 and 6.32~MeV correspond to the $p$ $1/2^-$ and $3/2^-$ hole
states, respectively.
The small amounts of the $p$~$3/2^-$ strength are also fragmented to the
states of 9.93~MeV and 10.7~MeV.

\begin{figure}[tbp]
\includegraphics[width=7.0cm]{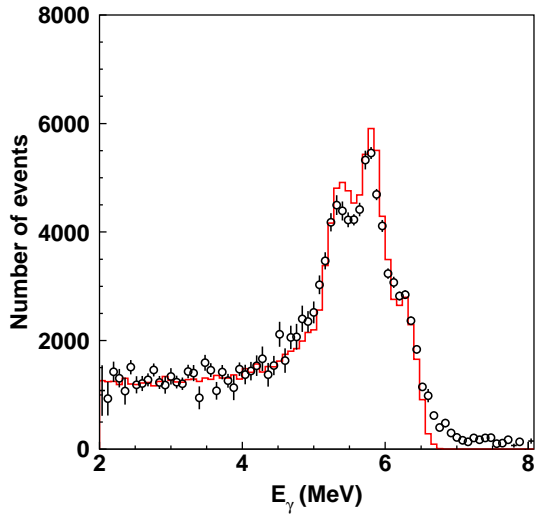}
\caption{\label{fig:p6mev} The coincidence $\gamma$-ray spectrum with the NaI scintillators obtained by gating on the peak at $E_x$=5.3--7.3~MeV in the \ox reaction. The open circles and histogram show the data and 6.32~MeV $\gamma$-ray MC, respectively.}
\end{figure}

In the analysis of the $\gamma$-rays from the hole states in $^{15}$N,
we use the independent signal of each NaI scintillator to determine the
$\gamma$-ray energy.
Since we take the timing data of the proton beam and NaI scintillator hit,
the accidental coincidence backgrounds are estimated by using the events
in the neighboring beam bunches of the true coincident timing.
Using the 6.32~MeV $\gamma$-ray from the $3/2^-$ state, the time variation of 
each PMT gain is monitored and corrected for. The 6.32~MeV $\gamma$-ray data 
is obtained by gating on $E_x$=5.3--7.3~MeV as shown in Fig.~\ref{fig:p6mev}.
In the following analysis, to estimate the $\gamma$-ray emission probabilities 
from the $s$-hole state in $^{15}$N, we generate $\gamma$-ray Monte Carlo 
(MC) simulations. 
We use GEANT 3 package~\cite{mc} for simulation of particle tracking.
The energy resolutions of the $\gamma$-ray detectors are taken into 
account from the 6.32~MeV $\gamma$-ray data.

\subsection{\label{sec:sstate}$\gamma$-rays with $E_{\gamma}<10$~MeV }
\begin{figure}[tbp]
\includegraphics[width=8.cm]{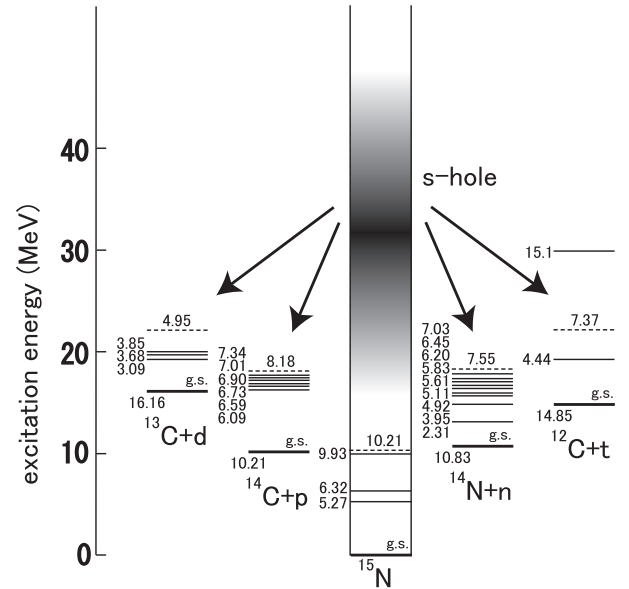}
\caption{\label{fig:scheme} A decay scheme from the $s$-hole state in
 $^{15}$N~\cite{scheme}. The bold solid lines show ground states.
 The narrow solid lines show all the possible excited states to emit
 de-excitation $\gamma$-rays below particle emission thresholds 
 (the break lines), except for the 15.1~MeV state. The states in
 $^{13}$C are also fed by $p$+$n$~decay, and the states in $^{12}$C are
 fed by $d$+$n$~decay and $p$+$n$+$n$~decay. We do not show the
 $^{13}$N+$n$+$n$~decay since the two-neutron emission threshold is high (21.39~MeV).}
\end{figure}

Figure~\ref{fig:scheme} shows the decay scheme of the $s$-hole state 
in $^{15}$N.
From the $s$-hole state in $^{15}$N, different particles are emitted.
In Ref.~\cite{yosoi}, the yields of these emitted particles were measured 
with their energies.
The fraction of the $\alpha$ emission was found to be very small and excited
levels fed by proton, neutron, deuteron, and triton emissions were 
observed.
Some of the residual nuclei remained in an excited state below the
particle re-emission threshold.
This indicates that secondary $\gamma$-rays are emitted.
\begin{table}
\begin{center}
\caption{\label{tab:fitstate} Candidate states to be generated by a particle emission from the $s$-hole state in $^{15}$N~\cite{scheme}. The emission probabilities for the $\gamma$-rays with $E_{\gamma}>3$~MeV are shown. $N(k)/N_{\rm tot}$ are obtained from the fitting. The numbers in the parentheses in the $N(k)/N_{\rm tot}$ are the sums of $N(k)/N_{\rm tot}$ at the 7.01~MeV state in $^{14}$C and the 7.03~MeV state in $^{14}$N and at the 6.09, 6.59, and 6.90~MeV states in $^{14}$C, respectively.}
\begin{tabular}{cD..{1.2}lD..{1.9}D..{1.2}}
\hline
\hline
 & \multicolumn{2}{c}{Energy level} & \multicolumn{1}{c}{$\gamma$-ray energy} & \multicolumn{1}{c}{}\\
Decay scheme & \multicolumn{2}{c}{MeV (J$^{\pi}$)} & \multicolumn{1}{c}{MeV (ratio)} & \multicolumn{1}{c}{$N(k)/N_{\rm tot}$}\\
\hline
\hline
 $^{13}$C+$d$ & 3.09 &(1/2$^+$) & 3.09\ (100\%)& 3.0\%\\
\hline
$^{13}$C+$d$ & 3.68 &(3/2$^+$) & 3.68\ (99.3\%)& 4.2\%\\
\hline
$^{13}$C+$d$ & 3.85 &(5/2$^+$) & 3.09\ (1.20\%)& 4.6\%\\
 & & & 3.68\  (36.3\%)&\\
 & & & 3.85\  (62.5\%)&\\
\hline
$^{12}$C+$t$ & 4.44 & (2$^+$) & 4.44\ (100\%)& 5.8\%\\
\hline
$^{14}$N+$n$ & 4.92 &(0$^-$) & 4.92\ (97\%)& 5.2\%\\
\hline
$^{14}$N+$n$ & 5.11 &(2$^-$) & 5.11\ (79.9\%)& 0.0\%\\
\hline
$^{14}$N+$n$ & 5.69 &(1$^-$) & 3.38\ (63.9\%)& 4.5\%\\
 & & & 5.69\  (36.1\%)& \\
\hline
$^{14}$N+$n$ & 5.83 &(3$^-$) & 5.11\ (62.9\%)& 0.54\%\\
 & & & 5.83\  (21.3\%)&\\
\hline
$^{14}$N+$n$ & 6.20 &(1$^+$) & 3.89\ (76.9\%)& 0.0\%\\
 & & & 6.20\  (23.1\%)&\\
\hline
$^{14}$N+$n$ & 6.45 &(3$^+$) & 5.11\ (8.1\%)& 2.8\%\\
 & & & 6.44\  (70.1\%)&\\
\hline
$^{14}$N+$n$ & 7.03 &(2$^+$) & 7.03\ (98.6\%)&(6.7\%)\\
\hline
$^{14}$C+$p$ & 6.09 &(1$^-$) & 6.09\ (100\%)& (0.0\%)\\
\hline
$^{14}$C+$p$ & 6.59 &(0$^+$) & 6.09\ (98.9\%)& (0.0\%)\\
\hline
$^{14}$C+$p$ & 6.73 &(3$^-$) & 6.09\ (3.6\%)& 0.43\%\\
 & & & 6.73\  (96.4\%)&\\
\hline
$^{14}$C+$p$ & 6.90 &(0$^-$) & 6.09\ (100\%)& (0.0\%)\\
\hline
$^{14}$C+$p$ & 7.01 &(2$^+$) & 6.09\ (1.4\%)&(6.7\%)\\
 & & & 7.01\  (98.6\%)&\\
\hline
$^{14}$C+$p$ & 7.34 &(2$^-$) & 6.09\ (49.0\%)& 5.7\%\\
 & & & 6.73\ (34.3\%)&\\
 & & & 7.34\ (16.7\%)&\\
\hline
\hline
\end{tabular}
\end{center}
\end{table}

\begin{figure*}[tbp]
\includegraphics[width=5.7cm]{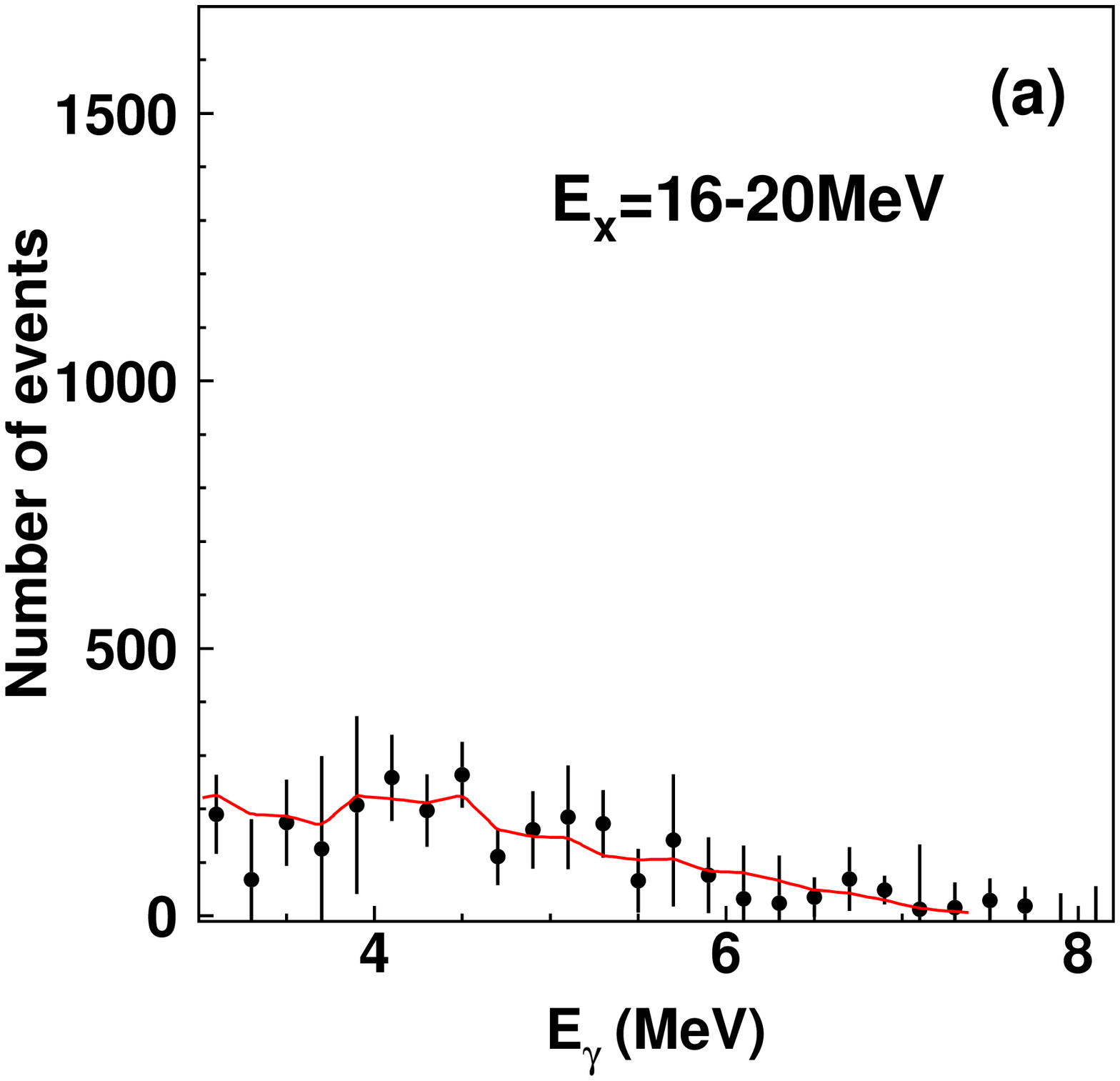}
\includegraphics[width=5.7cm]{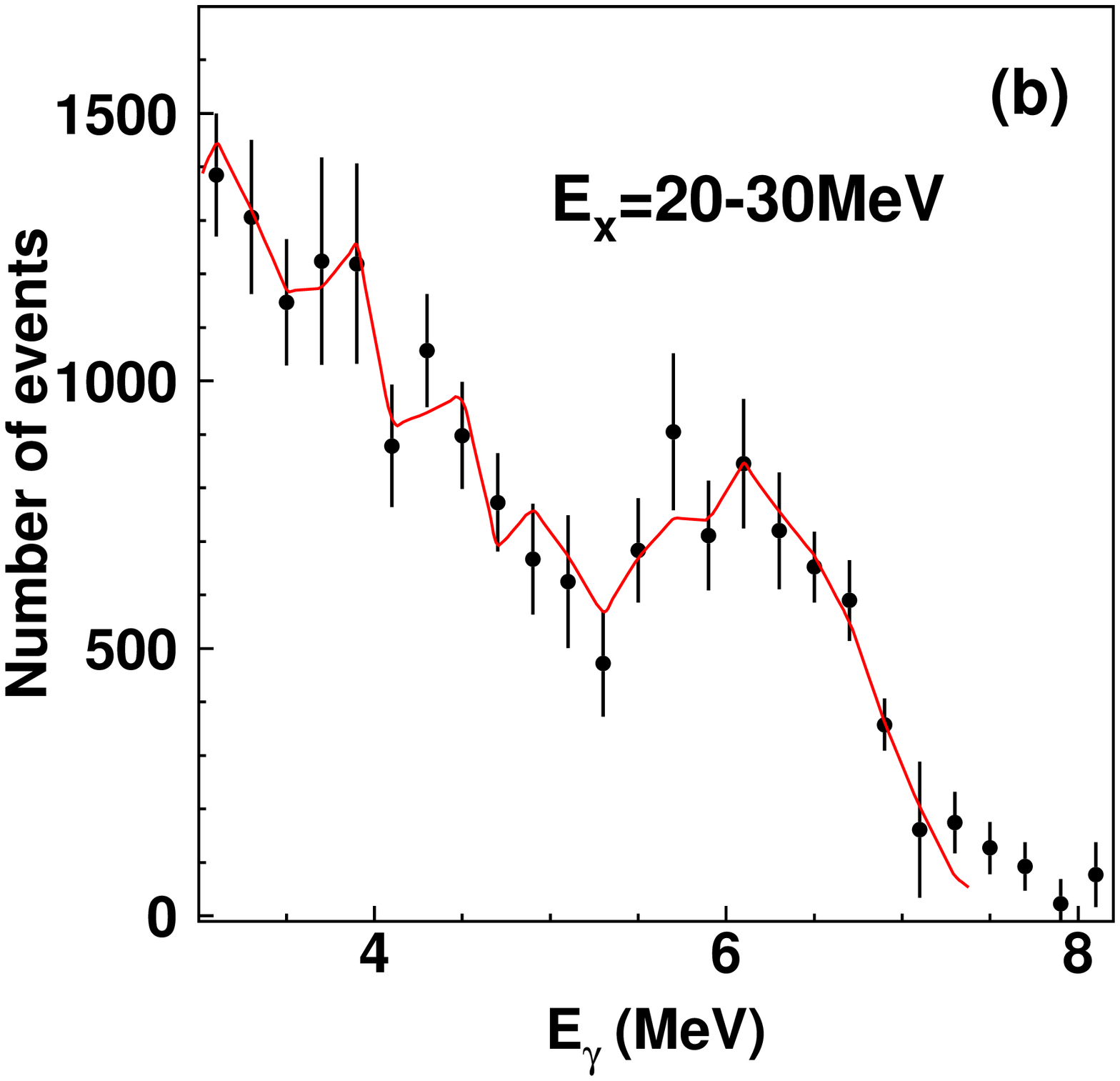}
\includegraphics[width=5.7cm]{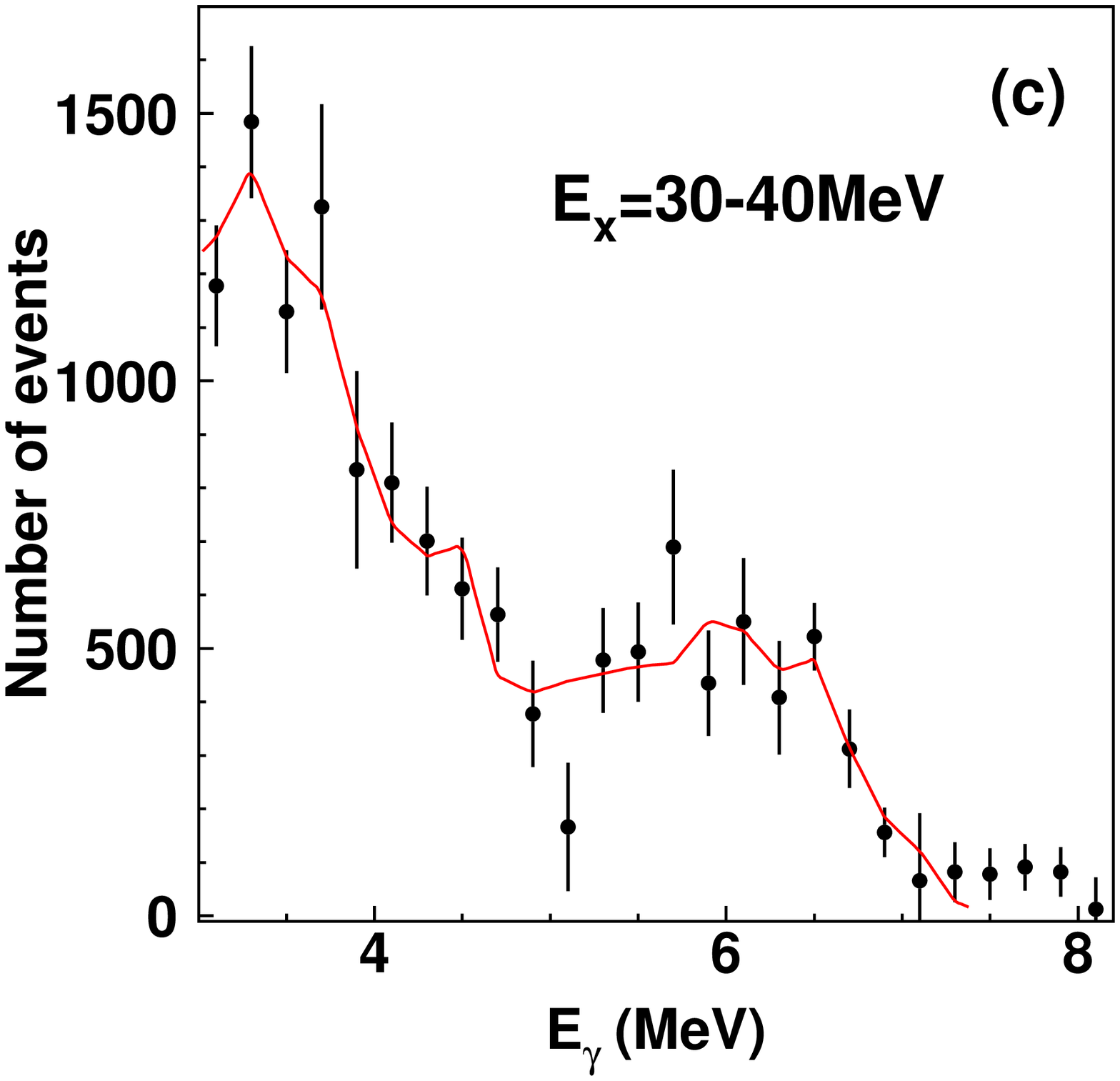}
\caption{\label{fig:shole} The energy spectrum of $\gamma$-rays from the $s$-hole state in $^{15}$N after subtracting accidental coincidence background. Each figure is obtained in coincidence by gating two proton events in the excitation-energy regions at (a) $E_x$=16--20~MeV, (b) $E_x$=20--30~MeV, and (c) $E_x$=30--40~MeV via the \ox reaction. The closed circles and lines show the data and the results of the best fitted MC simulations, respectively.}
\end{figure*}

The $\gamma$-ray data from the $s$-hole state are obtained by gating the
two-proton events in the excitation-energy region at $E_x$=16-40~MeV.
In the higher excitation-energy region, the signal-to-noise ratio becomes 
worse, because the detection efficiencies decrease gradually due to the 
finite momentum acceptance of the spectrometers.
Therefore, we do not use the data with $E_x>$40~MeV.
The $s$-hole state has sub-structures as mentioned above.
To specify $\gamma$-ray emissions from each sub-structure, the data is 
divided into three regions, 
$E_x$=16--20~MeV, $E_x$=20--30~MeV, and $E_x$=30--40~MeV,
which are indicated as A,B, and C in Fig.~\ref{fig:miss} (b), respectively.
Figure~\ref{fig:shole} shows the $\gamma$-ray energy distributions obtained
by gating the two-proton events in these three excitation-energy regions.
High energy $\gamma$-rays which are within the detectable range in \sk 
are clearly observed in Fig.~\ref{fig:shole} (b) and (c).

In order to obtain accurate values for these $\gamma$-ray emission 
probabilities, we fit the data with the associated $\gamma$-ray MC simulations.
We use the $\gamma$-ray data $E_{\gamma}$=3.0--7.4~MeV.
Because many kinds of $\gamma$-ray energy are associated,
we do not analyze data at $E_{\gamma}<$3.0~MeV.
An upper limit on the $\gamma$-ray gate energy of $E_{\gamma}<$7.4~MeV
is chosen because the highest excitation energy of the associated states 
below the particle emission thresholds is 7.34~MeV, as shown in 
Fig.~\ref{fig:scheme}.

From the excited states in Fig.~\ref{fig:scheme}, we choose 
candidate excited states that emit the $\gamma$-rays as listed in 
Table~\ref{tab:fitstate} for fitting. 
We omit the 3.95~MeV state in $^{14}$N because it 
mostly de-excites with $\gamma$-rays with $E_{\gamma}<$3.0~MeV.
Since the energies of two $\gamma$-rays with $E_{\gamma}$=7.01~MeV
and 7.03~MeV are very close, we treat these two $\gamma$-rays as 
mono-energetic.
We generate sixteen $\gamma$-rays in the MC simulations for seventeen
states in total.
We then estimate the yield of the states by fitting the $\gamma$-ray data
with these MC simulations.
The result of the best fitted MC simulations are shown in Fig.~\ref{fig:shole}
(a), (b), and (c) in comparison with the data.
The fitted lines reproduce the data well.

From the fitting result, the branching ratio to the state $k$ 
($N(k)/N_{\rm tot}$) is evaluated, where $N(k)$ is the number of events 
at the state $k$ and $N_{\rm tot}$ (=2.49$\times 10^{6}$) is the number 
of two-proton events in the excitation-energy region at $E_x$=16--40~MeV 
in $^{15}$N.
These branching ratios are summarized in Table~\ref{tab:fitstate}.
Since Cherenkov light is generated from all decay $\gamma$-rays,
the total de-excitation energy of $\gamma$~decay ($E_{\gamma,{\rm tot}}$)
determines the amount of Cherenkov light. 
These states de-excite only by emitting one or more $\gamma$-rays.
Therefore, $N(k)/N_{\rm tot}$ corresponds to the $\gamma$-ray emission 
branching ratio with the total energy of $E_{\gamma,{\rm tot}}$ equivalent to 
the excitation energy of the state $k$. 
In \sk, $\gamma$-rays with $E_{\gamma,{\rm tot}}>6$~MeV are well within 
the detectable energy range for the \nukp analysis, and $\gamma$-rays with 
$E_{\gamma,{\rm tot}}$=3--6~MeV are in the partially detectable range.
Therefore, we estimate the summed branching ratios of the $\gamma$-ray 
emission with $E_{\gamma,{\rm tot}}>6$~MeV and 
$E_{\gamma,{\rm tot}}$=3--6~MeV, and also estimate the systematic 
uncertainties of $\sum_k N(k)$ as summarized in Table~\ref{tab:syserr}.
We take into account the following systematic uncertainties: imperfect 
knowledge of $\gamma$-ray energy scale, detector acceptance, and 
selection of the fitting states.
From the 0.2~MeV error of the energy scale, we estimated
the uncertainty to be $^{+1.6}_{-5.4}$\% and $^{+9.9}_{-3.4}$\% for
$E_{\gamma,{\rm tot}}>6$~MeV and $E_{\gamma,{\rm tot}}$=3--6~MeV, 
respectively.
The uncertainty of the detector acceptance is estimated to be 3.4\%
using the 6.32~MeV $\gamma$-ray from the $3/2^-$ $p$-hole state, because 
its decay branching ratio is 100\%.
To check the effect of the selection of the fitting states, we apply the same
fitting method omitting any one of the fourteen states.
We take the largest difference from the original result as the uncertainty.
The summed emission branching ratios ($\sum_k (N(k)/N_{\rm tot})$)
with their systematic uncertainties are estimated to be \resbra for 
$\gamma$-rays with $E_{\gamma,{\rm tot}}>6$~MeV and 
$E_{\gamma,{\rm tot}}$=3--6~MeV, respectively.
This means that the large amount (about a half) of the $s$-hole state
decays to the excited states of daughter nuclei after the particle
emissions, which agrees with the results of particle-decay 
measurement~\cite{yosoi} and supports the theoretical prediction by the 
SU(3) model~\cite{yamada}.

{\renewcommand\arraystretch{1.4}
\begin{table}
\begin{center}
\caption{\label{tab:syserr} Summary of systematic uncertainties (\%) of $\sum_k N(k)$ with $E_{\gamma,{\rm tot}}>6$~MeV and $E_{\gamma,{\rm tot}}$=3--6~MeV.}
\begin{tabular}{ccc}
\hline
\hline
 & $E_{\gamma,{\rm tot}}>6$~MeV & $E_{\gamma,{\rm tot}}$=3--6~MeV\\
\hline
energy scale & $^{+1.6}_{-5.4}$\% & $^{+9.9}_{-3.4}$\% \\
detector acceptance & $\pm$3.4\% & $\pm$3.4\% \\
state selection & $^{+0.90}_{-0.72}$\% & $^{+5.8}_{-8.0}$\% \\
\hline
 total & $^{+3.9}_{-6.4}$\% & $^{+12.}_{-9.3}$\% \\
\hline
\hline
\end{tabular}
\end{center}
\end{table}
}
We evaluate the total $\gamma$-ray emission 
probabilities from the $s$-hole state associated with proton decay
in $^{16}$O.
The emission probabilities ($P(k)$) are evaluated using the equation
$P(k) = (S_p(s{\rm -state})/8)\times(N(k)/N_{\rm tot})$.
The spectroscopic factor of the $s$-hole state in the excitation-energy
region at $E_x$=16--40~MeV is estimated to be 1.6 using the 
distorted wave impulse approximation (DWIA) calculation~\cite{yosoi,yosoi3}.
The uncertainty of the spectroscopic factor is estimated to be about 10\% 
from the DWIA calculations using different optical potentials.
The total $\gamma$-ray emission probabilities with 
$E_{\gamma,{\rm tot}}>6$~MeV and $E_{\gamma,{\rm tot}}$=3--6~MeV are 
estimated to be \respro, respectively.
Because some extra $s$-hole strengths still exist at $E_x >$ 40~MeV, the total 
emission probability for the whole $s$-hole state become slightly larger.
In addition to the simple direct knockout ($p,2p$) reaction, the multi-step
processes or correlated processes like ($p,3p$) and ($p,2pn$) reactions
contribute the excitation spectrum.
$N_{\rm tot}$ includes such non-quasi-free contributions.
The difference between $(N(k)/N_{\rm tot})_{s-hole}$ and $(N(k)/N_{\rm tot})$
is estimated to be 15\% for the extreme case that $N(k)$ 
is zero for the non-quasi-free process.
The proton decay is a slightly different process from the ($p,2p$) reaction,
because the nuclear medium could affect the proton decay and not leave
a simple hole state. The fraction of the correlated decay is estimated to be
10\% by Ref.~\cite{yamazaki}. In this paper, we evaluate the emission 
probabilities without this effect.

The $\gamma$-ray emission probabilities estimated for the 7.01 and 7.03~MeV
states are less than half of Ejiri's estimate~\cite{ejiri}.
However, we found a few excited states to emit $\gamma$-rays 
with $E_{\gamma}>6$~MeV.
Compared to the 4\% $\gamma$-ray emission probability with 
$E_{\gamma}>6$~MeV estimated by Ejiri,
our emission probabilities are estimated to be 3.1\%.
On the basis of the $(e,e^{\prime}p)$ result~\cite{leus}, the dominant 
6.32~MeV $\gamma$-ray from the  $3/2^-$ $p$-hole state is expected to 
be emitted with 29\% 
($P(6.32{\rm \ MeV\ } p_{3/2^-})=S_p(6.32{\rm \ MeV\ }p_{3/2^-})/8=2.35/8$). 
The emission probabilities of 3.1\% with $E_{\gamma,{\rm tot}}>6$~MeV 
is found to be about 1/10 compared with that of the 6.32~MeV $\gamma$-ray.
Moreover, we found that the emission probability with 
$E_{\gamma,{\rm tot}}$=3--6~MeV is high, though no $\gamma$-rays are quoted
in this energy region in Ref.~\cite{ejiri}.
Especially, the 4.4~MeV $\gamma$-ray emission mainly from the 
$^{12}$C+$t$~decay is strong, despite the high $Q$-value.
This is consistent with the result of the particle-decay 
measurement~\cite{yosoi}.
The reason why the triton decay probability is higher than that of
$\alpha$~decay is theoretically explained by the selection rule 
obtained from the spatial SU(3) symmetry~\cite{yamada}.

\subsection{\label{sec:15mev}15.1MeV $\gamma$-ray}
\begin{figure}[tbp]
\includegraphics[width=7.0cm]{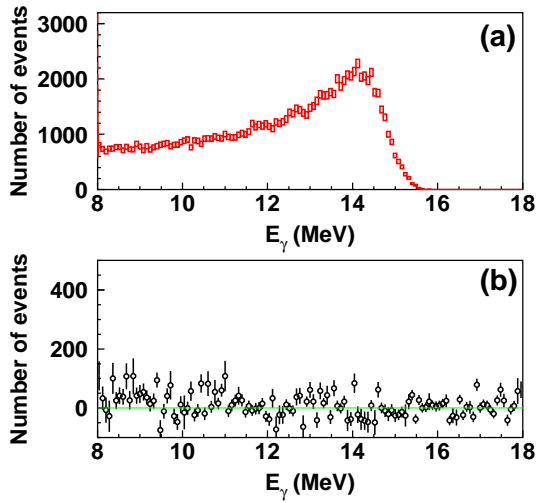}
\caption{\label{fig:15mev} The $\gamma$-ray energy distribution obtained by gating two-proton events in coincidence in the excitation-energy region at $E_x$=16--40~MeV via the \ox reaction. The MC simulations of the 15.1~MeV $\gamma$-ray (a) and data with statistical errors (b) are shown, respectively.}
\end{figure}

A 15.1~MeV $\gamma$-ray is emitted mainly from the T=1,~$1^+$ state in 
$^{12}$C which might be made from the $s$-hole state in $^{15}$N$^*$.
In the $\gamma$-ray data obtained by gating on the $s$-hole events in 
the excitation-energy region at $E_x$=16--40~MeV,
we search for a signal in the energy region at $E_{\gamma}$=10--15.7~MeV,
where the peak is expected to be located based on the MC study.
Figures~\ref{fig:15mev} (a) and (b) show the 15.1~MeV $\gamma$-ray MC 
simulation and data, respectively.
We do not find any significant excess within the statistical uncertainty.
The limit on the total emission probability including the $s$-hole strength
of the 15.1~MeV $\gamma$-ray is estimated to be 0.38\% at 99\% confidence 
level.

The branching ratio of the door-way $s$-hole state in $^{15}$N to the 
T=1,~1$^+$ state in $^{12}$C has been recently calculated in the same
manner as used in Ref.~\cite{yamada2}. 
The obtained value is about 0.04\%~\cite{c15theo}.
Taking into account the 15.1~MeV emission branching ratio from T=1,~$1^+$ 
state in $^{12}$C and the spectroscopic factor of the $s$-hole state in 
$^{15}$N, the total emission probability is predicted to be $\sim$0.007\%.
Although the statistical decay process from the $s$-hole state might
contribute to produce the 15.1~MeV state in $^{12}$C, much more
statistics of the data is needed to find out the signal of 15.1~MeV
$\gamma$-ray.

\section{\label{sec:concl}Conclusion}
We studied the de-excitation $\gamma$-rays from the excitation of the 
$s$-hole state in $^{15}$N via the \ox reaction.
The emission branching ratio of $\gamma$-rays with more-than-6~MeV and 
3--6~MeV from the $s$-hole state in the excitation-energy region of 16 
to 40~MeV are estimated to be \resbra, respectively. 
If we take into account the spectroscopic factor of the $s$-hole state,
the total emission probabilities are found to be \respro, respectively.
In water Cherenkov detector experiments, it is important to understand the
decay process with $\gamma$-rays from the proton-hole state in $^{15}$N.
Especially for the proton decay search via \nukp in \sk, this result is 
useful to reduce the systematic uncertainty of the detection efficiency.
Moreover, we searched for the 15.1~MeV $\gamma$-ray from the $s$-hole state.
However, we do not find any signal. We estimate the upper limit on 
the emission probability to be 0.38\% at 99\% confidence level.

\section{\label{sec:ack}Acknowledgments}
We are grateful to the RCNP cyclotron staff for preparing a stable and clean 
beam.
This research was supported in part by the Grant-in-Aid for Scientific 
Research Nos.~07404010 and 09440105 from the Japan Ministry of Education,
Sports, Culture, Science, and Technology.


\end{document}